\documentclass{IOS-Book-Article}

\usepackage{mathptmx}

\usepackage{amsmath,amssymb,amsfonts}
\usepackage{graphicx,psfrag,epsf}
\usepackage{enumerate}
\usepackage{url} 
\usepackage{tikz}
\usepackage{pgf}
\usepackage{setspace}

\newtheorem{prop}{Proposition}

\def\hb{\hbox to 10.7 cm{}}

\def\inde{\hbox{$\perp$\hskip-6.5pt$\perp$}}

\begin{document}

\def\thepage{}

\begin{frontmatter}              

\setcounter{page}{157}

\title{On Associative Confounder Bias}


\author[A]{\fnms{Priyantha} \snm{Wijayatunga}%
\thanks{Corresponding Author: Priyantha Wijayatunga, Department of Statistics, Ume\r{a} School of Business and Economics,  Ume\r{a} University, Ume\r{a} SE-901 87, Sweden. Email: priyantha.wijayatunga@umu.se.}},
\runningauthor{P. Wijayatunga}
\address[A]{Department of Statistics, Ume\r{a} University, Ume\r{a}, Sweden}

\begin{abstract}
Conditioning on some set of confounders that causally affect both treatment and outcome variables can be sufficient for eliminating bias introduced by all such confounders when estimating causal effect of the treatment on the outcome from observational data. It is done by including them in propensity score model in so-called potential outcome framework for causal inference whereas in causal graphical modeling framework usual conditioning on them is done.  However in the former framework, it is confusing when modeler finds a variable that is non-causally associated with both the treatment and the outcome. Some argue that such variables should also be included  in the analysis for removing bias. But others argue that they introduce no bias so they should be excluded and conditioning on them introduces spurious dependence between the treatment and the outcome, thus resulting extra bias in the estimation. We show that there may be errors in both the arguments in different contexts. When such a variable is found neither of the actions may give the correct causal effect estimate. Selecting one action over the other is needed in order to be less wrong. We discuss  how to select the better action.
\end{abstract}

 
\pagestyle{empty}
\thispagestyle{empty}

\end{frontmatter}


\section{Introduction} \label{sec:intro}

For making causal inferences from observational data (see \cite{RD2005} and \cite{PJ2009}) it is important to find, ideally all the potential pretreatment confounders of the given causal relation between the cause (treatment variable) and the effect (outcome variable), in order to obtain unbiased causal effect estimate of the former on the latter. Let $Z$ denote the treatment received by subjects, taking values from the set $\mathcal{Z}=\{0,1 \}$ and let $Y$ denote the outcome, taking values from the set $\mathcal{Y}=\{0,1\}$ where $0$ denotes failure and $1$ denotes success. In potential outcome causal model \cite{RD2005} it is accepted  existence of pair of potential outcomes $(Y_1, Y_0)$ for each subject, where $Y_i$ is the outcome that would have been observed had the treatment been $Z=i$ for $i=0,1$. It is assumed that the pair is independent of the treatment assignment, written as $(Y_0,Y_1) \inde Z $ when the treatment assignments are randomized as in case of a randomized experiment.  However in observational studies, the treatment assignments are not randomized. Then, useful assumption for causal inference is that the potential outcomes are conditionally independent of the treatment assignment given the pretreatment covariates, say, multivariate $X$.  Ideally, $X$ denotes 'all' the potential pretreatment confounders of $Z$ and $Y$ and then it is written as $(Y_1,Y_2) \inde Z \mid X$. That is, to estimate the causal effect of $Z$ on $Y$, we need to condition on (control for) $X$. However, it is not necessary to consider all the pretreatment confounders but any 'sufficient' subset of them. Finding such a sufficient set of condounders is somewhat problematic and the potential outcome framework offers no clear way to do it. However, causal graphical modeling framework \cite{PJ2009} offers one way that is called 'back door criterion'. It shows how to choose a subset of covariates in order to identify the causal effect (to estimate it without bias).   When a causal graphical model is identified on $Z$, $Y$ and all their causal factors the criterion can find a sufficient subset and such a set is called  an 'admissible' or a 'deconfounding' set in the literature. Considering some covariates as confounders by ignoring such a criterion can sometimes introduce   further bias (p. 351 of \cite{PJ2009}). However, the back-door criterion is not complete \cite{PJ2009}; there exist causal graphical models where the criterion fails for some sets of covariates though adjusting for them results in valid causal effect estimates.

So, the problem of confounder selection is important in casual inference. In  the potential outcome causal model, when the analyst has found  all the confounders then he/she uses  them either directly or indirectly (in  so-called propensity score models \cite{RR1983}, \cite{RR1984}) for removing induced bias  from them. However, any factor that is causing both the treatment and the outcome could be identified relatively easily as pretreatment confounders  with subject domain knowledge. For that it is important to decide causal directions among the variables. But there may be other factors such as the ones that are non-causally related with both the treatment and the outcome, for e.g., those with associations. It seems that some researchers tend to use them for conditioning too, for e.g., including them in the propensity score model assuming that it removes the bias due to them.  However generally, the causal graphical modelers do not consider  them as confounders. Recently there was a debate (see \cite{SI09} \cite{PJ09a}, \cite{SA09} and  \cite{RD09}) on this issue; if it is necessary to condition on a variable that is not causally related with both  the treatment and the outcome but associated with both. In the debate, Rubin argues for and Pearl and his colleagues argue against saying that it will only introduce extra bias. Our goal here is to analyze these arguments a little more deeper and to understand  when we should condition on them. We use graphical modeling framework to estimate causal effects therefore, we begin by giving some details of it. We argue that in some cases, it is desirable to condition whereas in others, it is not.  Mostly the decision should be taken considering strengths of associations of the potential confounder with the treatment and the outcome.  

\section{Covariate Selection for Adjustment of Confounding}

We use concept of intervention in causal graphical models (also called do-calculus) described in \cite{PJ2009} and \cite{LR2002} for the causal effect estimation. This approach is equivalent to the potential outcome model (see Ch. 7 of \cite{PJ2009} and \cite{WP2014}). To recall the reader with this calculus, first define the probability distribution of a random variable with conditioning by intervention or action on another variable.  For an observed random data sample on a vector of random variables, say, $\textbf{X}=(X_1,...,X_n)$, we can find the joint probability distribution of them, say, $p(\textbf{X}=\textbf{x})=p(\textbf{x})$. We can have a factorization of $p(\textbf{x})$; let it be  $p(\textbf{x})= \prod_{i}^{n} p(x_i \vert pa_i)$ where $PA_i \subseteq \{X_1,...,X_{i-1}$\} with the exception of $PA_1= \emptyset$ (empty set) using some conditional independence assumptions within $\textbf{X}$. Note that here we denote random variables (or sets of them) by uppercase letters/expressions (such as $X, PA$, etc.) and their values by relevant lowercase expressions ($x,  pa$, respectively). For a causal structure on $\textbf{X}$  one can use, for e.g., time order of happening to index the variables such that cause variables have higher indices than those of effect variables'. For any $i$, such that $2 \leq i \leq n$ if $p(x_i \vert pa_i) \neq p(x_i )$ then the probability distribution of vector of random variables without $X_i$, say, $\textbf{X}_{-i}=\{X_1,...,X_n\} \backslash \{X_i\}$ when $X_i$  is intervened to a particular value of it, say, $x_i$, written as $do(X_i=x_i)$, denoted by $p(\textbf{x}_{-i} \vert do(X_i=x_i))$ is defined as follows; 
\begin{eqnarray*}
p(\textbf{x}_{-i} \vert do(X_i=x_i)) &=& \frac{p(\textbf{x})}{p(x_i \vert pa_i)}=\prod_{k=1:k \neq i}^{n} p(x_k \vert pa_k) \\
 & \neq & \frac{p(\textbf{x})} {p(x_i )} =\frac{1}{p(x_i)}\prod_{k=1}^{n} p(x_k \vert pa_k) = p(\textbf{x}_{-i} \vert x_i) 
\end{eqnarray*}
where the last expression is corresponding conditional probability distribution when we have observed $X_i=x_i$. That is, generally two probability distributions differ.  

\small
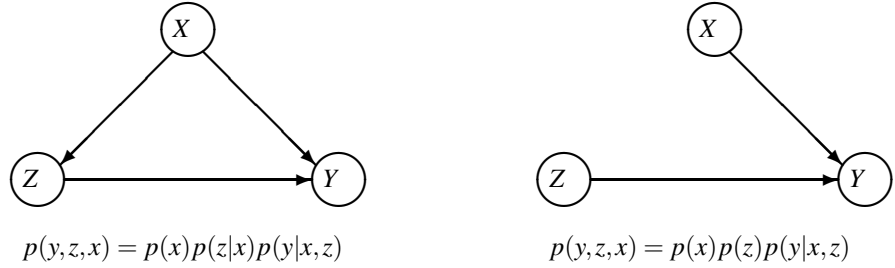
\begin{figure} 
\begin{center}
\setlength{\unitlength}{1mm}
\begin{picture}(100,40)(-50,-10)
\thicklines
\put(-30,20){\circle{7}}
\put(-32,19){$X$}

\put(-50,0){\circle{7}}
\put(-52,-1){$Z$}

\put(-10,0){\circle{7}}
\put(-12,-1){$Y$}

 \put(-32,17){\vector(-1,-1){15}}
 \put(-28,17){\vector(1,-1){15}}
 \put(-46.5,0){\vector(1,0){33}}
\put(-52, -10){$p(y,z,x)=p(x)p(z\vert x)p(y\vert x,z)$}

\put(40,20){\circle{7}}
\put(38,19){$X$}

\put(20,0){\circle{7}}
\put(18,-1){$Z$}

\put(60,0){\circle{7}}
\put(58,-1){$Y$}

 \put(42,17){\vector(1,-1){15}}
 \put(23.5,0){\vector(1,0){33}}

\put(18, -10){$p(y,z,x)=p(x)p(z)p(y\vert x,z)$}

\end{picture}
\end{center}
\caption{ \label{simple.bn}  Two Bayesian network causal models}
\end{figure}
\normalsize 
Now, consider two different causal relationships between $X$, $Y$ and $Z$: the first one is such that $X$ is a cause of both $Z$ and $Y$, and $Z$ is a cause of $Y$ which is  represented as  causal network model $p(y,z,x)=p(x)p(z\vert x)p(y\vert x,z)$ shown by left hand side diagram and the second one is such that $X$ and $Z$ are causes of $Y$ which is  represented as a causal network model $p(y,z,x)=p(x)p(z)p(y\vert x,z)$ shown by right  hand side diagram in the Figure \ref{simple.bn}. And if we intervene on $Z$ as $do(Z=z)$ for $z=0,1$, then marginal intervention distribution of $Y$ for the  first causal model  is  $p(y \vert do(Z=z)) = \sum_x p(x)p(z \vert x)p(y \vert z,x)/p(z \vert x) = \sum_x p(y \vert z,x)p(x)$ whereas that for the second causal model is $ p(y\vert do(Z=z) )= \sum_x p(x)p(z)p(y \vert z,x)/p(z) = \sum_x p(y \vert z,x)p(x) = p(y \vert z) $, since $X \inde Z$ in latter case. And the causal effect of the treatment option $Z=z_1$ compared to the control option $Z=z_0$ is defined as $\sum_y y p(y \vert do(Z=z_1)) - \sum_y y p(y \vert  do(Z=z_0))$. It is identifiable if $\sum_x p(y \vert z,x)p(x)$ is a valid functional for $z=z_0,z_1$.  Then we see that the estimates for the two cases are different.

\begin{figure}
\begin{center}
\setlength{\unitlength}{1mm}
\begin{picture}(100,40)(-50,-10)
\thicklines

\put(-30,20){\circle{7}}
\put(-32,19){$X$}

\put(-50,0){\circle{7}}
\put(-52,-1){$Z$}

\put(-10,0){\circle{7}}
\put(-12,-1){$Y$}

\put(-50,20){\circle{7}}
\put(-51,19){$U$}

\put(-10,20){\circle{7}}
\put(-12,19){$W$}

 \put(-32,17){\vector(-1,-1){15}}
 \put(-28,17){\vector(1,-1){15}}
 \put(-46.5,0){\vector(1,0){33}}

\put(-46.5,20){\vector(1,0){13}}
\put(-50,16.5){\vector(0,-1){13}}

\put(-13.5,20){\vector(-1,0){13}}
\put(-10,16.5){\vector(0,-1){13}}

\put(30,20){\circle{7}}
\put(28,19){$X$}

\put(10,0){\circle{7}}
\put(8,-1){$Z$}

\put(50,0){\circle{7}}
\put(48,-1){$Y$}

 \put(28,17){\vector(-1,-1){15}}
 \put(32,17){\vector(1,-1){15}}
 \put(13.5,0){\vector(1,0){33}}

\put(10,20){\circle{7}}
\put(9,19){$U$}

\put(50,20){\circle{7}}
\put(48,19){$W$}

 \put(10,16.5){\vector(0,-1){13}}
 \put(13.5,20){\vector(1,0){13}}

 \put(50,17){\vector(0,-1){13}}
 \put(46.5,20){\vector(-1,0){13}}

\qbezier(10,24)(30,30)(50,24)

 \put(13.5,0){\vector(1,0){33}}

\put(-35,-10){Model 1}

\put(25,-10){Model 2}


\end{picture}
\end{center}
\caption{ \label{simple.bn2}  Model 2 is obtained by extending Model 1}
\end{figure}
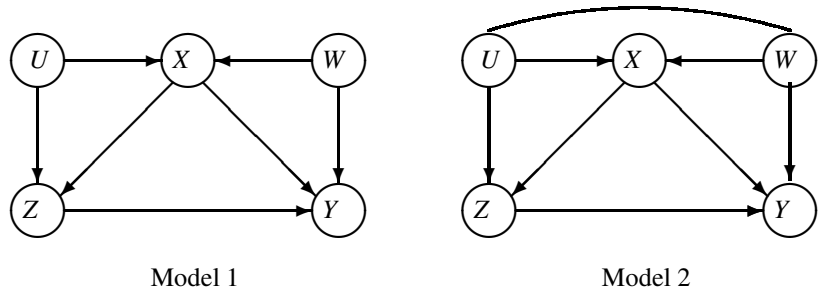

The above observation can be shown for a more general causal model. Let $\textbf{X}=(X_1,...,X_n)$ be according to time order and $\textbf{X}_0$ represent a set variables that causally affect $X_p$ but we are not sure about chronological order of the elements of $\textbf{X}_0$ with $X_i$ for $1 \le i < p \le n$.  Let parents (causes) of $X_j$ in $\textbf{X}$ be $PA_j$ and that in $(\textbf{X}_0,\textbf{X})$ be $PA_j^+$, so $PA_j=PA_j^+$ for $j \leq p-1$ and $PA_p^+=PA_p \cup \textbf{X}_0$.  Then, the joint probability distribution of $(\textbf{X},\textbf{X}_0)$ is $p(\textbf{x}_0, \textbf{x}) = p(\textbf{x}_0) \prod_{j=1}^{n} p(x_j \vert pa_j^+)$ and the intervention (on $X_i$) distribution is  
\begin{eqnarray*}
p( x_p \vert do(X_i=x_i))  &=& \sum_{\substack{\textbf{x}_0,x_1,..x_{i-1}, \\ x_{i+1},...,x_{p-1}, \\  x_{p+1},...,x_{n}}} p(\textbf{x}_0) \prod_{\substack{j=1 \\ j \neq i }}^{n} p(x_j \vert pa_j^+)  \\
                                     &=& \sum_{\substack{\textbf{x}_0,x_1,..x_{i-1}, \\ x_{i+1},...,x_{p-1}}} p(\textbf{x}_0) \prod_{\substack{j=1 \\ j \neq i }}^{p-1} p(x_j \vert D_j) p(x_p \vert D_p,\textbf{x}_0) \\
                                    &=& \sum_{\substack{\textbf{x}_0,x_1,..x_{i-1}, \\ x_{i+1},...,x_{p-1}}} p(\textbf{x}_0) p(x_1,..x_{i-1}) p(x_{i+1},....,x_p \vert D_{i+1},\textbf{x}_0) \\
                                     &=& \sum_{\textbf{x}_0,pa_p} p(\textbf{x}_0)p(pa_p \backslash \{ x_i\}) p(x_p \vert pa_p,\textbf{x}_0) \\
                                  &=& \sum_{\textbf{x}_0,pa_p} p(x_p \vert x_i,pa_p \backslash \{ x_i\},\textbf{x}_0)p(pa_p \backslash \{ x_i\},\textbf{x}_0)
\end{eqnarray*}
where $D_j=\{ X_1,...,X_{j-1}\}$ such that $D_1=\emptyset$ and $X_i \in PA_p$.  This is of the form of $p(y \vert do(z))=\sum_x p(y \vert z,x)p(x)$ where $Z$ and $X$  affects $Y$ directly and so is $X$ on $Z$. If we assume that some of the variables in $\textbf{X}_0$ are associated with some of the variables in the vector $(X_1,...,X_{i-1})$ or, causally related or associated with variables in $(X_{p+1},...,X_n)$ then the above result holds.  
\begin{eqnarray*}
p( x_p \vert do(X_i=x_i))  &=& \sum_{\textbf{x}_0,pa_p} \frac{ p(x_p, x_i,pa_p \backslash \{ x_i\},\textbf{x}_0)}{ p(x_i, pa_p \backslash \{ x_i\},\textbf{x}_0) / p(pa_p \backslash \{ x_i\},\textbf{x}_0)}  \\
                                     &=& \sum_{\textbf{x}_0,pa_p} \frac{ p(x_p, x_i,pa_p \backslash \{ x_i\},\textbf{x}_0)}{ p(x_i \vert pa_p \backslash \{ x_i\},\textbf{x}_0) } = \sum_{\textbf{x}_0,pa_p} \frac{ p(x_p, x_i,pa_p \backslash \{ x_i\},\textbf{x}_0)}{ p(x_i \vert pa_p^-) }  \\
                                    &=&  \sum_{\textbf{x}_0,pa_p^-} \frac{ p(x_p, x_i,pa_p^-)}{ p(x_i \vert pa_p^-) }  =  \sum_{\textbf{x}_0,pa_p^-} p(x_p \vert x_i,pa_p^-)p( pa_p^-)
\end{eqnarray*}
where $PA_p^- = PA_p \cap PA_i$. Again, this is in the form of $p(y \vert do(z))=\sum_x p(y \vert z,x)p(x)$ where $X$ represents all the direct causal variables common to both $Z$ and $Y$. And above simplifications show that we can select the confounding variable set as follows. 
\begin{prop} Let $X^{\prime \prime}$ denote the set of all potential causal variables of $Y$ except for $Z$ and let $P(Y \vert Z,X^{\prime \prime})=P(Y \vert Z,X^{\prime})$ where $X^{\prime}$ is the smallest subset of $X^{\prime \prime}$ in some sense. Then the smallest subset $X$ of $X^{\prime}$ in a similar sense such that $P(Z \vert X^{\prime})=P(Z \vert X)$ is a sufficient set of confounders for estimating $P(Y \vert do(Z))$.
\end{prop}
Here the smallest subset $A$ of $X$ can be a set of variables whose  sum of their configurations is the smallest. This rule gives a simple way to select covariates for removing confounding bias.  We avoid the proof of this rule but it is clear from the above discussion. Recall that the back-door criterion is known to be incomplete (see Ch. 11 of  \cite{PJ2009})  meaning that the criterion fails for some sets of covariates but adjusting for them is sufficient for removing confounding bias. Above rule avoids inclusion of covariates such as instrumental variables, especially for building propensity score models. In fact, in literature  sufficient confounder set is selected such that, firstly each confounder in it is a cause of the treatment, and then it is a cause of the outcome \cite{VS2011}.  However, it should be done in the other way round; a confounder should be predictive of the outcome first and then it should also predictive of the treatment. Following this order we do not miss any important confounders, since any confounder should be related to the outcome at the first place.  For e.g., consider a causal model for  estimating causal effect of teacher's instructional practice ($Z$) on student's reading comprehension achievement ($Y$) as discussed in \cite{KC2011}. It is assumed that the  teacher's reading knowledge $X$ is a causal confounder such that it  affects directly both $Z$ and $Y$. Furthermore, it is assumed that the teacher's professional development in reading $U$ affects directly to $Z$ and $X$ and the teacher's general knowledge ($W$) affects directly to $X$ and $Y$. The causal diagram is shown as Model 1 in Figure~\ref{simple.bn2}.  Then it is easy to see that $p(y \vert do(z))=\sum_x p(y \vert z,x)p(x)$. And if we believe that $U$ and $W$ are dependent, for e.g., through a common cause, then we get that $p(y \vert do(z))=\sum_{x,w} p(y \vert z,x,w)p(x,w)$ when $Z \not\inde W$, that is reasonable to assume. 

\small
\normalsize

\section{Associative Confounders}

There is a controversy among the research community about kinds of  variables that should be considered as confounders for including, especially in the propensity score models in the potential outcome causal model, since therein the causal diagrams showing the causal structure are often not used. In fact, initially the propensity score concept came into light to describe the treatment allocation process \cite{RR1984}, \cite{RR1983}. In the current practice some authors argue that all the variables related to outcome should be included in the propensity score model \cite{RD1996} (there can be some redundancy then) whereas others argue that all the variables related to both the treatment and the outcome should be included \cite{PS2000}. However, the problems occur when one finds variables that have non-causal (associative) relationships with the treatment or the outcome.   Researchers usually replace any such association between two variables with a causal fork using so-called common cause principle. This is to replace an association with causal relations \cite{SA09}.  Simply, the principle says that a non-causal association between two variables can be replaced by a third variable that is causally affecting the both. For e.g., such an association between two variables $X$ and $Y$ with a model, say, $M_1 : X \rule[1.3mm]{4mm}{0.1pt} Y $ can be replaced by a model, say,  $M_2: X \leftarrow U \rightarrow Y$ where arrows indicate causal relations. Then, $U$ is said to be a common cause of $X$ and $Y$. Note that we omit the possibility of having feedback causal relations.      

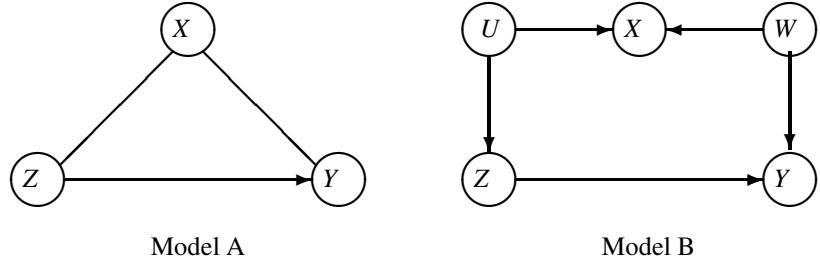
\begin{figure}
\begin{center}
\setlength{\unitlength}{1mm}
\begin{picture}(100,40)(-50,-10)
\thicklines

\put(-30,20){\circle{7}}
\put(-32,19){$X$}

\put(-50,0){\circle{7}}
\put(-52,-1){$Z$}

\put(-10,0){\circle{7}}
\put(-12,-1){$Y$}

 \put(-32,17){\line(-1,-1){15}}
 \put(-28,17){\line(1,-1){15}}
 \put(-46.5,0){\vector(1,0){33}}

\put(30,20){\circle{7}}
\put(28,19){$X$}

\put(10,0){\circle{7}}
\put(8,-1){$Z$}

\put(50,0){\circle{7}}
\put(48,-1){$Y$}

 \put(13.5,0){\vector(1,0){33}}

\put(10,20){\circle{7}}
\put(9,19){$U$}

\put(50,20){\circle{7}}
\put(48,19){$W$}

 \put(10,16.5){\vector(0,-1){13}}
 \put(13.5,20){\vector(1,0){13}}

 \put(50,17){\vector(0,-1){13}}
 \put(46.5,20){\vector(-1,0){13}}

 \put(13.5,0){\vector(1,0){33}}

\put(-35,-10){Model A}

\put(25,-10){Model B}


\end{picture}
\end{center}
\caption{ \label{causalfork.bn}  Model B is obtained by applying two separate causal forks to Model A}
\end{figure}

Now, let we observe a covariate $X$ that is non-causally associated with both $Z$ and $Y$, which is the topic of Rubin and Pearl debate. It can be assumed that the non-causal association structures  $Z \rule[1.3mm]{4mm}{0.1pt} X  \rule[1.3mm]{4mm}{0.1pt} Y$, is embedded in the context and therefore,  apply a causal fork to each of the two associations separately.  In fact, the argument of \cite{SA09} and    \cite{PJ09a} is based on applying two causal forks for the two non-causal associations, one for each, thus  making $X$ a so-called M-collider \cite{KC2011}.  Their model of discussion is the Model A in Figure \ref{causalfork.bn} but the argument is based on the model $ Z \leftarrow U \rightarrow  X \leftarrow W \rightarrow Y$ that is called an M-structure due to its shape. Here $U$ and $W$ are taken to be independent.  An example of this model is given in \cite{DL2010}: measuring causal effect of low education ($Z$) on later diabetes risk ($Y$) where it is assumed that mother's previous diabetes status ($X$) is an associative covariate. A medical opinion is that family income during the childhood ($U$) is a cause of $X$ and $Z$, and mother's genetic risk of diabetes ($W$) is a cause of $X$ and $Y$.  

Though $U$ and $W$ can be independent, it is appropriate to think  that it is a special case and generally, there is  some dependence between them. In fact, one can just assume it but here we investigate how and when such cases arise and  discuss which actions are appropriate then. For Model B of Figure $\ref{causalfork.bn} $, we can write the joint probability distribution of all the variables as  $ p(y,x,u,w,z) = p(u)p(w)p(x \mid u,w)p(z \vert u)p(y\vert z,w)$, so with intervention $Z=do(z)$, we get $p(y,x,u,w \vert do(z)) = p(u)p(w)p(x \vert u,w)p(y \vert z,w) $. Then,
\begin{align*}
p(y \vert do(z)) &= \sum_{u,w,x} p(u)p(w)p(x \vert u,w)p(y\vert z,w) = \sum_{w} p( w)p(y\vert  z,w)   = \sum_{w} p( w \vert z)p(y\vert z,w)          \\
                       &= p(y \vert z) = \sum_x p(y \mid z,x) p(x \mid z)  \neq  \sum_x p(y \mid z,x)p(x)
\end{align*}
if $W \inde Z$ and,  since $Z \not\inde X$. Note that we have $W \inde Z$ whenever $U \inde W$. That is, the true probability of $Y$ when $Z$ is intervened is different from that obtained by conditioning on $X$.  And ignoring $X$ gives the true intervened probability. So, when assuming $U \inde W$, conditioning on $X$ may result in a biased causal effect estimate; above inequality shows that the biasness may have caused due to the dependence between $X$ and $Z$, since $p(x) \neq p(x \vert z)$, i.e., when $X$ and $Z$ are weakly dependent the biasness is small. Note that if an error is occurred in the estimate of $p(y \vert do(z))$ for $z=0,1$ then it may not necessarily result in an error of same magnitude  in causal effect estimate that is $\sum_y yp(y \vert do(Z=1)) -\sum_y yp(y \vert do(Z=0))$, i.e., two errors may result in a different error. Resultant error (bias) due to conditioning on $X$ is  $\sum_{y,x} y p(y \vert Z=1,x) [p(x) - p(x \vert Z=1)] - \sum_{y,x} y p(y \vert Z=0,x) [p(x) - p(x \vert Z=0)] $. For simplicity, we concentrate on errors that can occur in estimation of $p(y \vert do(z))$ for $z=0,1$. Note that, above discussion is valid when some of other confounders, say, $X_1$ are present where $X \inde X_1$. And in the above analysis we made a strong assumption that $W \inde Z$, but this may not sometimes be true in reality.  In the following section we show this possibility.

\small
\normalsize

\subsection{Dependence of $X$ with $Z$ and $Y$}

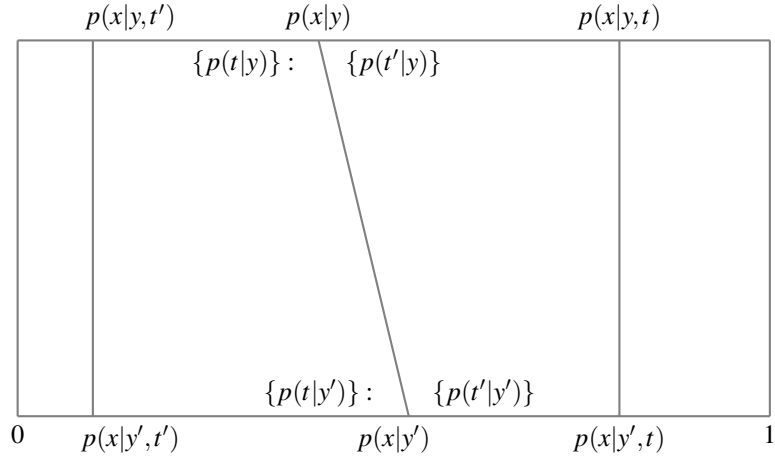
\begin{figure}[!ht]
\begin{center}
\begin{tikzpicture}[scale=10]
    \draw [thick, gray, -] (0,0) -- (0,0.5);      

    \draw [thick, gray, -] (0,0) -- (1,0);      


    \node [below] at (1,0) {$1$};               
    \node [below] at (0,0) {$0$};               

\draw [thick, gray, -] (0,0.5) -- (1,0.5);
\draw [thick, gray, -] (1,0) -- (1,0.5);
\draw [thick, gray, -] (0.1,0) -- (0.1,0.5);
\node [below] at (0.15,0) {$p(x \vert y',t')$};
\node [below] at (0.8,0) {$p(x \vert y',t)$};

\draw [thick, gray, -] (0.8,0) -- (0.8,0.5);
\node [above] at (0.15,0.5) {$p(x \vert y,t')$};
\node [above] at (0.8,0.5) {$p(x \vert y,t)$};

\draw [thick, gray, -] (0.52,0) -- (0.4,0.5);
\node [below] at (0.5,0) {$p(x \vert y')$};
\node [above] at (0.4,0.5) {$p(x \vert y)$};

\node [above] at (0.4,0) {$ \{ p(t \vert y') \}: $};
\node [above] at (0.62,0) {$ \{p(t' \vert y') \} $};

\node [below] at (0.3,0.5) {$ \{ p(t \vert y) \}: $};
\node [below] at (0.5,0.5) {$ \{p(t' \vert y) \} $};
\end{tikzpicture}
\caption{A common cause variable $T$ for the negative correlation between $X$ and $Y$, $p(x \vert y) < p(x \vert y')$. For the probabilities $p$ and $q$ where $p+q=1$ the expression $\{p\}$:$\{q\}$ means that the lengths of two line segments on which $p$ and $q$ appear are according to the ratio $p:q$. \label{fig:fig1}}
\end{center}
\end{figure}

It is natural to consider the cases when $W  \not\inde Z$ then it may be that $p(y \vert do(z)) \neq p(y \vert z)$ even if $U \not\inde W$. However, since $W$ is hidden it is  unclear how to consider this case. In fact, for Model B in Figure~\ref{causalfork.bn} we have $W \inde Z$ \cite{LS1990}. Let us assume the case that $X$ and $Y$ are strongly dependent. We use   a geometric figure that is used to visualize the Simpson's paradox \cite{WP2015a} to explore this possibility. Let the association between $Y$ and $X$ be such that $p(x \vert y) < p(x \vert y')$. Then, for some $T$ we have that $p(t' \vert y)p(x \vert y,t')+p(t \vert y)p(x \vert y,t) < p(t' \vert y')p(x \vert y',t')+p(t \vert y')p(x \vert y',t)$. Note that there are infinitely many such $T$ but they can be artificial unless given some meaningful interpretation, ideally to few of them.   Now consider the case of $p(x \vert y,t')< p(x \vert y,t)$. It is important to note that the value $p(x \mid y)$ dissects  positive length $p(x \vert y,t)-p(x \vert y,t')$ according to ratio $p(t \vert y):p(t' \vert y)$;
\begin{align*}
\{p(t \mid y)+p(t' \mid y)\}p(x \mid y)&=p(t' \mid y)p(x \mid y,t') + p(t \mid y)p(x \mid y,t) \\
p(t \mid y) \{p(x \mid y,t)-p(x \mid y) \} &= p(t' \mid y) \{p(x \mid y)-p(x \mid y,t') \} \\
\frac{p(x \mid y)-p(x \mid y,t')}{p(x \mid y,t)-p(x \mid y)} &=\frac{p(t \mid y)}{p(t' \mid y)}
\end{align*}
Now if $T$ is a common cause of $X$ and $Y$ association then we should have $p(x \vert y,t)=p(x \vert y',t)=p(x \vert t)$  and $p(x \vert y,t')=p(x \vert y',t')=p(x \vert t')$. Therefore, in  Figure~\ref{fig:fig1} the conditional probabilities in the former equality are vertically aligned, and so are those in the latter. Then we have $p(x \vert y',t')< p(x \vert y',t)$ and $p(x \vert y')$ dissects  positive length $p(x \vert y',t)-p(x \vert y',t')$ according to ratio $p(t \vert y'):p(t' \vert y')$ and similarly for $p(x \vert y,t')< p(x \vert y,t)$ and $p(x \vert y)$.   In the Figure~\ref{fig:fig1}  those ratios are marked  with braces. Since the selection of  $T$ is restricted by the strength of the dependence between $X$ and $Y$, for a higher value of it, we can have a higher dependence between $Y$ and $T$. And if the strength of the dependence between $X$ and $Y$ is characterized by $p(y \vert x)$ then that between $X$ and $T$ should be also higher.  And the other case is similar, i.e., taking $p(x \vert y,t') > p(x \vert y,t)$.

\begin{figure} 
\begin{center}
\setlength{\unitlength}{1mm}
\begin{picture}(100,50)(-50,-10)

\thicklines

\put(-30,15){\circle{7}}
\put(-32,14){$X$}

\put(-50,0){\circle{7}}
\put(-52,-1){$Z$}

\put(-10,0){\circle{7}}
\put(-12,-1){$Y$}

 \put(-46.5,0){\vector(1,0){33}}

\put(-30,30){\circle{7}}
\put(-32,29){$V$}

\put(-33,28){\vector(-2,-3){16}}
\put(-27,28){\vector(2,-3){16}}
\put(-30,26.5){\vector(0,-1){8}}

\put(-35,-10){Model C}

\put(30,15){\circle{7}}
\put(28,14){$X$}

\put(10,0){\circle{7}}
\put(8,-1){$Z$}

\put(50,0){\circle{7}}
\put(48,-1){$Y$}

 \put(13.5,0){\vector(1,0){33}}

\put(10,20){\circle{7}}
\put(9,19){$U$}

\put(50,20){\circle{7}}
\put(48,19){$W$}

 \put(10,16.5){\vector(0,-1){13}}
 \put(13.5,19){\vector(4,-1){13}}

 \put(50,16.5){\vector(0,-1){13}}
 \put(46.5,19){\vector(-4,-1){13}}

\put(13.5,21){\vector(1,0){33}}

\put(13.5,0){\vector(1,0){33}}

\put(25,-10){Model D}

\end{picture}
\end{center}
\caption{ \label{complexcon.bn}  Common cause models}
\end{figure}
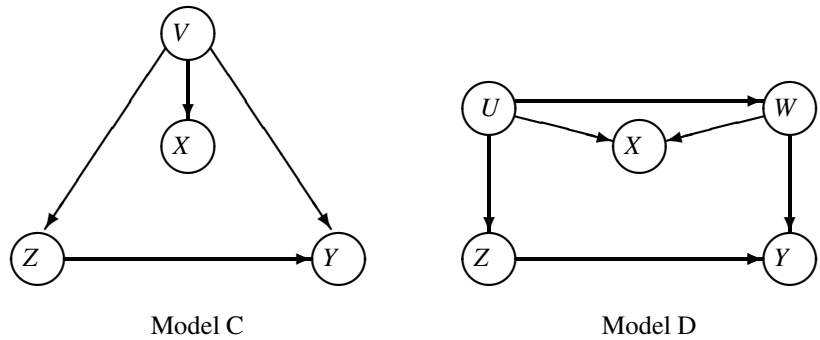

If $T$ is $W$ in our causal model in Figure~\ref{causalfork.bn}, a common cause for the association between $X$ and $Y$, then the dependences between $X$ and $W$, and $Y$ and $W$ can be strong given that the dependence between $X$ and $Y$ is strong. Similarly, a strong association between $Z$ and $X$ implies that those between $Z$ and $U$, and  $U$ and $X$ can be strong. With similar arguments, these imply that $U$ and $W$ can be dependent. An alternative way to see that $U$ and $W$ are not independent when the associations between $X$ and $Z$, and $X$ and $Y$ are strong is to use correlations. In \cite{LE2001}  it is shown that for any three random variables, say, $A,B$ and $C$ the correlation coefficients among them satisfy the relationship $\rho_{AC}^2+\rho_{BC}^2+\rho_{AB}^2 \leq 1+2\rho_{AC}\rho_{BC }\rho_{AB}$.  If, for e.g., when $\rho_{XZ}=0.8$ and $\rho_{XY}=0.7$ then we cannot have $U$ and $W$ such that $\rho_{UW}=0$. Therefore, when the dependences between $Z$ and $X$,  and $Y$ and $X$ are strong it may be that the introduced two common causes for those associations are dependent. Furthermore, there can be another possibility for these two associations; both associations may be due one cause, i.e., both $U$ and $W$ refer to the same hidden variable ($V$ in the Model C in Figure~\ref{complexcon.bn}). 

However, current studies are often done without considering these possibilities. But some researchers have shown that conditioning on associated covariates introduces only a small bias.  Their claims may be due to these contexts. Sometimes it is advised \cite{RD09} to control for all the pretreatment covariates  but the graphical causal model researchers reject this idea.  Therefore, in the next section we take a look at different possibilities of associative covariates and try to understand when the biasness can be amplified.

\small
\normalsize

\subsection{Deciding on Conditioning}

Consider the case of two dependent hidden causes, i.e., $U \not \inde W$ (such as Model D in Figure \ref{complexcon.bn}) where the dependence is causal or non-causal. Then,
\begin{align*}
p(y \vert do(z)) &= \sum_{u, w, x} p(u,w)p(x \vert u,w)p(y \vert z,w)=\sum_{w} p(w)p(y \vert z,w) = \sum_{w,x} p(x)p(w \vert x)p(y \vert w,z,x) \\ 
                       & \neq \sum_{w,x} p(x) p(w \vert z,x)p(y \vert w,z,x) = \sum_x p(y \vert z,x)p(x)                
\end{align*}
if $p(w \vert x) \neq p(w \vert z,x )$ i.e., $W \not\inde Z \vert X$, conditioning on $X$ does not give the correct probability estimate that is $\sum_{w} p(y \vert z,w)p(w)$.  And ignoring $X$ also does not give the correct estimate, since then we get  $p(y \vert do(z)) = p(y \vert z) =\sum_w p(y \vert z,w)p(w \vert z) \neq \sum_{w} p(y \vert z,w)p(w)$, i.e., we need to assume $Z \inde W$ in order to have the correct probability for the case but we know that $Z \not\inde W$, especially when associations between $X$ and $Y$, and $Z$ and $Y$ are strong. That is, to condition on $X$ we should have $W \inde Z \vert X$ and to  ignore $X$ we should have $W \inde Z$.   So, the question remains is that which statement should be accepted in order to be more correct against the other; either $W \inde Z  \vert X$ or  $W \inde Z$. Accepting the former (rejecting the latter) is to condition on $X$ and vice versa. But none of the conditions can be tested, since they involve unobservable $W$. 

 However, with some subject domain knowledge if one can assume meaningful $U$ and $W$ and then recognize their dependences with $X$ (based on those between $Z$ and $X$, and $Y$ and $X$) it may be possible to decide which option can be better. For e.g., if those dependences are not strong and causation of $U$ and $W$ on $X$ is mostly based on explaining away phenomenon \cite{PJ1988}, then it may not be desirable to condition of $X$. Note that the explaining away phenomenon is that when we see $X=1$ then observing $U=1$ makes $P(W=1)$ lower and vice versa.  If conditioned on $X$ in this case, then comparative  strata of data sample in the causal effect calculation may have imbalances in the causal variables $U$ and $W$. This can cause biased causal effect estimates. And when the dependences of $U$ and $W$ with $X$ is assumed to be high then it is less likely that there is an explaining away phenomenon, i.e, most probably the causation is monotonic (when we see $X=1$ then observing $U=1$ makes $P(W=1)$ higher and vice versa) then conditioning on $X$ can be beneficial because it results in balances in the causal variables $U$ and $W$. Though one can reason about the actions to be taken as done above, it requires extensive simulation studies to confirm them.  

Now consider the case of single hidden cause, say, $V$ (Model C in Figure \ref{complexcon.bn}). Then
\begin{align*}
p(y \vert do(z)) &= \sum_{x,v} p(v)p(x \vert v)p(y \vert z,v)=\sum_{v} p(y \vert z,v) p(v)= \sum_{x,v} p(x)p(v \vert x)p(y \vert z,v,x) \\
                       & \neq \sum_{x,v} p(x)p(v \vert z,x)p(y \vert z,v,x) = \sum_x p(y \vert z,x)p(x).                
\end{align*}
Therefore, here also conditioning on $X$ does not gives the correct probability estimate that is $\sum_{v} p(y \vert z,v) p(v)$ if $p(v \vert x) \neq p(v \vert z,x)$, i.e., $V \not \inde Z \vert X$.  But ignoring $X$ also does not give the correct estimate as $p(y \vert do(z)) \neq p(y \vert z)$ in this case. Since $ p(y \vert z) =\sum_v p(y \vert z,v)p(v \vert z)$, ignoring $X$ means assuming $V \inde Z$, but we know that $Z$ and $V$ should be  dependent. So, similar to the above case, the question remains is that which should be accepted against the other in order to be more correct; either or $V \inde Z  \vert X$ or  $V \inde Z$. Accepting the former is to condition on $X$ and vice versa. But similar to the above case where the dependences are higher, assuming $V \inde Z \vert X$ can be better  than assuming $V \inde Z $, therefore conditioning on $X$. If the subject domain knowledge shows that there is a single common cause $V$ then it is beneficial to condition on $X$.

\section{Conclusion}

Causal effect estimation tasks from observational data need to consider confounders of the causal relation of interest for controlling for (conditioning on). However, it is not necessary that all of them are considered but a "sufficient" subset of them. Often the current practice is to select them according to their predictive ability of the treatment firstly and then the outcome. But it should be done other way round; firstly they should be predictive of the outcome and then the treatment. And we show how to handle associative confounders (those are not causing both the treatment and outcome but associated with them) where currently there is no clear consensus about using them. It is often beneficial to condition on associative confounders when they are strongly dependent with both the treatment and outcome whereas it is not so for weakly dependent ones. 

\paragraph{\textbf{Acknowledgments:} Financial support for this research is from the Swedish Research Council for Health, Working Life and Welfare (FORTE) and SIMSAM at Ume\r{a}. And the author is thankful to Slawomir Nowaczyk and anonymous referees for their comments.}


\begin{thebibliography}{99}

\bibitem{RD2005}
D. Rubin. Causal Inference Using Potential Outcomes: Design, Modeling, Decisions.
\textit{Journal of the American Statistical Association} \textbf{100}(469) (2005), 322--331.

\bibitem{PJ2009} 
J. Pearl, \textit{Causality: Models, Reasoning, and Inference},  Cambridge University Press, New York, 2009.

\bibitem{RR1983}  
P. R. Rosenbaum and D. B. Rubin, The central role of the propensity score in observational studies for causal effects.
\textit{Biometrika} \textbf{70}(1) (1983), 41--55.

\bibitem{RR1984}
P. R. Rosenbaum and D. B. Rubin, Reducing Bias in Observational Studies Using Subclassification on the Propensity Score.
\textit{Journal of the American Statistical Association} \textbf{79}(387) (1984), 516--524.

\bibitem{SI09}
I. Shrier, Letter to the Editor, \textit{Statistics in Medicine} \textbf{28} (2009), 1315--1318.

\bibitem{PJ09a}
J. Pearl,  Letter to the Editor, \textit{Statistics in Medicine} \textbf{28} (2009), 1415--1416.

\bibitem{SA09}
A. Sj\"{o}lander, Letter to the Editor, \textit{Statistics in Medicine} \textbf{28} (2009), 1416--1420.

\bibitem{RD09}
D. Rubin, Author's Reply, \textit{Statistics in Medicine}\textbf{28} (2009), 1420--1423.


\bibitem{LR2002}
 S. L. Lauritzen  and T. S. Richardson,  Chain graph models and their causal interpretations. 
\textit{Journal of Royal Statistical Society, Series B} \textbf{64}(3) (2002), 321--361.

\bibitem{WP2014}
P. Wijayatunga, Causal Effect Estimation Methods.
\textit{Journal of Statistical and Econometric Methods} \textbf{3}(2) (2014), 153--170. 


\bibitem{VS2011}  
T. J. VanderWeele and I. Shpitser,  A New Criterion for Confounder Selection.
\textit{Biometrics} \textbf{67} (2011), 1406--1413.


\bibitem{KC2011}  
B. Kelcey  and J. Carlisle, The Threshold of Embedded M Collider Bias and Confounding Bias.
\textit{Society for Research on Eductaional Effectiveness Conference}, (2011). http://files.eric.ed.gov/fulltext/ED519118.pdf


\bibitem{RD1996}  
D. B. Rubin, Matching using estimated propensity scores: related theory and practice.
\textit{Biometrics} \textbf{52} (1996), 249--264.

\bibitem{PS2000}  
S. M. Perkins, W. Tu, M. G. Underhill, X. H. Zhou and M. D.  Murray, M. D. The use of propensity scores in pharmacoepidemiologic research.
\textit{Pharmacoepidemiol Drug Safety} \textbf{9} (2000), 93--101.


\bibitem{DL2010} 
L. Dallolio, R. Bellocco, L. Richiardi and M. P. Fantini, M.P. Using directed acyclic graphs to understand confounding in observational studies. 
\textit{Biomedical Statistics and Clinical Epidemiology} \textbf{3}(2) (2010), 89--96.


\bibitem{LS1990}
Lauritzen, S. L., Dawid, A. P., B. N. Larsen,  and H.-G. Leimer, Independent Properties of Directed Markov Fields
\textit{Networks}, \textbf{20} (1990), 491--505.

\bibitem{WP2015a}
P. Wijayatunga, Viewing Simpson's paradox.
\textit{Statistica \& Applicazioni}, \textbf{XII}(2) (2014), 225--235.

\bibitem{LE2001}
E. Langford, N. Schwertman and M. Owens, Is the property of being positively correlated transitive?
\textit{The American Statistician} \textbf{55}(4) (2001), 322--324. 

\bibitem{PJ1988}
J. Pearl, \textit{Probabilsitic Reasoning in Intelligent Systems: Networks of Plausible Inference} (Second Edition), Morgan Kauffmann, San Mateo, CA., 1988.


\end{thebibliography}
\end{document}